\documentclass[journal]{IEEEtran}

\usepackage[pdftex]{graphicx}

\usepackage{array}
\usepackage{float}
\usepackage{multirow}
\usepackage{bm}
\usepackage{mdwmath}
\usepackage{xcolor,soul,framed}
\usepackage{cite}
\usepackage{eqparbox}
\usepackage{url}
\usepackage{amsmath}
\usepackage{amssymb}
\usepackage{textcomp}
\usepackage{arydshln}
\usepackage{multirow}
\usepackage{bm}
\usepackage{tikz}
\usetikzlibrary{trees,decorations,shadows}
\tikzset{level 1/.style={sibling angle=45,level distance=4mm}}
\usetikzlibrary{arrows.meta}
\usetikzlibrary{calc}
\usetikzlibrary{math}
\usepackage{pdftexcmds}

\usepackage{pgfplots}
\usepackage{pgfplotstable}
\usepackage{amsmath}\usepackage{amssymb,amsthm}
\usepackage{amsfonts}

\usepackage [autostyle, english = american]{csquotes}
\usepackage{dblfloatfix}
\MakeOuterQuote{"}

\newcommand{\buses}{$\mathcal{N}$}
\newcommand{\busesm}{\mathcal{N}}

\newcommand{\meas}{$\mathcal{M}$}
\newcommand{\measm}{\mathcal{M}}

\makeatletter
\newcommand{\strequal}[2]{\pdf@strcmp{#1}{#2}==0}
\makeatother

\ifCLASSOPTIONcompsoc
    \usepackage[caption=false, font=normalsize, labelfont=sf, textfont=sf]{subfig}
\else
\usepackage[caption=false, font=footnotesize]{subfig}
\fi

\begin{document}

\bstctlcite{IEEEexample:BSTcontrol}
   \title{A Framework for Constrained Static State Estimation in Unbalanced Distribution Networks}

\author{Marta~Vanin,~\IEEEmembership{Graduate Student Member,~IEEE, }
 Tom~Van~Acker,~\IEEEmembership{Member,~IEEE,}
	Reinhilde~D'hulst, 
       and~Dirk~Van~Hertem,~\IEEEmembership{Senior Member,~IEEE}

\thanks{M. Vanin, T. Van Acker, and D. Van Hertem are with the Research Group ELECTA, Department of Electrical Engineering, KU Leuven, 3001 Heverlee, Belgium. 
Reinhilde D'hulst is with VITO, Boeretang 200, 3400 Mol, Belgium.
All authors are also with EnergyVille, Thor Park 8310, 3600 Genk, Belgium.
Corresponding author: marta.vanin@kuleuven.be
}
}

\maketitle

\begin{abstract}
State estimation plays a key role in the transition from the passive to the active operation of distribution systems, as it allows to monitor these networks and, successively, to perform control actions. However, designing state estimators for distribution systems carries a significant amount of challenges. This is due to the physical complexity of the networks, e.g., phase unbalance, and limited measurements. Furthermore, the features of the distribution system present significant local variations, e.g., voltage level and number and type of customers, which makes it hard to design a "one-size-fits-all" state estimator. 
The present paper introduces a unifying framework that allows to easily implement and compare diverse unbalanced static state estimation models. This is achieved by formulating state estimation as a general constrained optimization problem.
The advantages of this approach are described and supported by numerical illustration on a large set of distribution feeders.
The framework is also implemented and made available open-source. 

\end{abstract}



\section{Introduction}\label{sec:introduction}
\subsection{Background and Motivation}\label{ssec:background}

\IEEEPARstart{S}{tate} estimation (SE) is standard industrial practice in transmission systems since the 1970s, pioneered by the work of Schweppe et al.~\cite{schweppe13}.
This is not the case for distribution systems, mainly because of their standard "passive" management philosophy and the high cost of monitoring such networks, due to their large extension and the high number of directly connected customers, especially at the low voltage level.
The study of SE for distribution networks (DNs) only began in the 1990s~\cite{DSSE95, barankelley, Meliopoulos}, when it started to become evident that distributed generation and the electrification of transport and heating systems might change the nature of DNs. These technologies are changing the previously underutilized and predictable characteristics of DNs, increasing the risk of voltage and congestion issues but also introducing the potential to perform control actions through these inverter-based devices. Developing SE that suit the needs and features of DNs is the first step towards understanding the impact of these new technologies and their active management.  However, the direct translation of transmission systems SE techniques to the distribution level is not possible~\cite{towardsSG}, due to the different topology, the non-negligible phase unbalance~\cite{DSSE95, LV-SE-smartmeters} and the scarce measurement availability~\cite{bible}. Furthermore, the large size and the unbalance of DNs can give rise to computational challenges~\cite{dellagiustina}. Even though numerous distribution system state estimation (DSSE) studies are available in the literature, there is no general consensus on the most suitable design criteria~\cite{Majdoub}. This is partly due to the significant feature variations of DNs, such as 1) voltage level, 2) location, e.g., urban, rural, 3) customer type, e.g., house, small industry, 4) percentage of customers with PV panels, electric vehicles, smart meters, etc. Therefore, “one-size-fits-all” solutions might not exist and investigating DSSE methods is still an open research question. This paper presents a framework that allows to consistently explore different static SE models for unbalanced systems, facilitating the comparison of different methods from the literature and the development of customized SE paradigms.

\subsection{Literature review}\label{ssec:literature-review}

The traditional SE approach consists of minimizing the weighted sum of squares of measurement "residuals", and is called weighted least squares (WLS) estimation~\cite{bible}. The residuals are the difference between the measured and expected value of a variable. The availability of relatively easy and efficient (Gauss-)Newton algorithms to solve the WLS problem is one of the reasons behind its success~\cite{Rakpenthai}. (Weighted) least absolute value (WLAV) minimization is also a popular approach: despite presenting computational challenges w.r.t. WLS, it proves to be better at handling bad data~\cite{Gol2014}. 

However, the following features of DNs can lead to ill-conditioning SE problems, compromising the effectiveness of Gauss-Newton methods: 1) the presence of multiple measurement sources with different weights, 2) the lack of measurement redundancy, 3) the high R/X ratio and 4) the interconnection of long and short lines~\cite{yao_distribution_2019}. An example of the first issue are "zero-injection buses", which can cause computational issues if introduced as "virtual measurements", and are often preferably modelled as equality constraints~\cite{Clements, Lin}. The large number of zero-injection buses in DNs, especially in the low voltage, makes constrained SE particularly interesting. Inequality constraints can also be found, in both static~\cite{Kliokys, bible, Kumagai} and dynamic~\cite{EKF} SE examples, for different puropses: to address partially observable networks~\cite{Kumagai}, to provide indications on the direction of current or power flows~\cite{bible, Kliokys} or to include phasor measurement units~\cite{EKF}. 

While dynamic SE is a popular topic in transmission system studies, fewer publications address dynamic DSSE, as so far it does not seem to provide much improvement over WLS methods~\cite{Huang}, unless a large number of synchronized phasor measurement units (PMUs)~\cite{Carquex, Song} are available, which is unrealistic for most DNs, due to their high cost~\cite{MadaniPMU}. Therefore, the focus of this paper is on static SE. 


Constrained formulations of the static SE problem for unbalanced systems recently appeared in references~\cite{Dzafic, Majumdar, Picallo, Klauber, Gol}. However,~\cite{Dzafic, Majumdar, Picallo} resort to Newton-like algorithms to solve the SE, rather than specific optimization methods. Optimization methods are only applied in case of unusual settings or formulations:~\cite{Madani, Zhang} and~\cite{Klauber, Gol} explore semi-definite programming (SDP), second-order cone programming (SOCP) and linear programming (LP) formulations of the SE problem. The recent interest in relaxations of the otherwise non-convex SE problem has the aim of ensuring convergence even if the problem is ill-conditioned, while linearizations~\cite{Gol} are deployed to reduce the computational complexity.

References~\cite{Madani, Zhang,Klauber, Gol} resort to third-party solvers to solve their convex or linear SE. This is a countertrend compared to the traditional SE approach, where SE model and algorithm are jointly developed to achieve shorter computational times~\cite{bible}. To the best of the authors' knowledge, there are no examples in the literature where off-the-shelf solvers are used to address standard SE problems, based on the non-convex 'AC' power flow equations.

Nevertheless, SE can always be reduced to a constrained optimization problem, with the objective of minimizing a given criterion, e.g., WLS, and it is possible to extend the "user-made model plus off-the-shelf solvers" paradigm to address the full AC SE. The structure taken by the SE makes it a special case of optimal power flow (OPF) problem. While this might appear unnecessary, considering the existence of easy and fast algorithms to solve the classical WLS AC SE, it presents a number of advantages, discussed in Section~\ref{sec:scope}.




\subsection{Scope of the paper and contributions}\label{sec:scope}

The core of SE, leaving aside observability analysis and bad data detection methods, is made up of different "blocks": estimation criterion, measurements number and type, network model, power flow formulation and variable spaces, to name the main ones. Their multiple possible combinations made the number of DSSE-related publications increase fast and in a rather sparse manner, so that the need for numerous review papers was felt even in the last few years' time alone:~\cite{Dehghanpour, taskforce, Primadianto, WangOverview}. However, the rigorous comparison of the different proposed approaches and results is not only labour intensive, but often impossible due to lack of information. In almost every paper in the literature review, at least one of the following is not given: power profiles data, solver tolerance, solver start value, solver time, or network data.

Furthermore, as discussed in the introduction, the diversity of DNs features makes specific solutions only ideal in a specific subset of DNs, and it can be time consuming to investigate the many possibilities for each situation.

The present work attempts to overcome the above-described shortcomings, facilitating the consistent comparison of different DSSE models and the design of tailored solutions for different DNs. 

This is done through a unifying conceptual framework that consists of formulating SE as an OPF problem. The resulting constrained optimization-based DSSE serves as a "superproblem" that encapsulates the standard WLS AC SE and allows to extend upon it, including under-determined systems and (in)equality constraints. DSSE is formulated in a general manner, that allows to easily modify and recombine the different previously described "blocks". Moreover, modelling DSSE from an optimization standpoint allows to implement relaxations of any involved equation. For example, the WLAV criterion can be exactly relaxed to a tractable linear objective~\cite{Irving}.

This conceptual framework is implemented and made available as an open-source tool~\cite{DSSE.jl}, based on 'PowerModelsDistribution.jl'~\cite{PMD_PSCC}, and the mathematical programming toolbox JuMP~\cite{JuMP}. PowerModelsDistribution is a package that features multiple unbalanced power flow formulations: exact AC, relaxations and linear approximations, in different variable spaces, that can be integrated in the proposed framework. 

JuMP provides an interface between models and off-the-shelf optimization solvers. This allows the user to edit the model at any time, without the need to adapt the algorithm. Also, optimization solvers are topology agnostic: they can be applied to radial and meshed systems alike. The drawback of separating model and solver is that solution times are longer than with a customized algorithm, but computational results show acceptable solve time. The intention of this work is anyway not to provide fast SE algorithms, but a framework to facilitate the design process. If faster solution times are crucial, a customized algorithm can be developed afterwards, once the optimal design is chosen. 

To the best of the author's knowledge, there is presently no available software that allows to compare DSSE methods. The only similar implementation are PandaPower~\cite{PandaPower}, which allows to perform SE, but only with the WLS criterion and balanced equations, and the DNToolBox~\cite{DNToolBox}, which allows unbalanced WLS calculations, but shares only part of the code and data and is not designed to be easily extended upon.

The comparison between different model and ease of design have the potential to accelerate the implementation of DSSE in pilot projects and real-life settings. Finally, together with the open-source code, all the data used in this work are made available, with the intention of encouraging benchmarking. 

The advantages of using the proposed framework are summarized in the following list:
\begin{itemize}
    \item Flexibility: easy interchange of components, formulations, measurement types, criteria, solvers, etc.,
    \item Extensibility: adding models/components or variations on the DSSE problem is straightforward,
    \item Benchmarking: results, data and code are provided,
    \item General problem formulation that allows to include (in)equality constraints, under-determined systems and relaxations,
    \item The "user model plus off-the-shelf solver paradigm" allows to exploit advances from both sides. Solve times are reasonable,
    \item Topology agnosticism: system can be meshed or radial.
\end{itemize}

The rest of the paper is structured as follows. Section~\ref{sec:mathematical_model} presents the mathematical model of the DSSE-OPF, Section~\ref{sec:case_studies} presents a number of examples to showcase some functionalities and advantages of the presented framework, with numerical results. Finally, conclusions are drawn in Section~\ref{sec:conclusion}.





%
%
%
\section{Mathematical Model}\label{sec:mathematical_model}
\newcommand{\z}{$\mathbf{z}$}
\newcommand{\h}{$\mathbf{h}$}
\newcommand{\hhm}{$\mathbf{h}_m$}
\newcommand{\errv}{$\boldsymbol{\eta}$}
\newcommand{\vm}{$|U|$}
\newcommand{\va}{$\angle U$}
\newcommand{\vmjp}{$|U_{j,p}|$}
\newcommand{\vajp}{$\angle U_{j,p}$}
\newcommand{\vi}{$U^{\text{im}}$}
\newcommand{\vr}{$U^{\text{re}}$}
\newcommand{\ca}{$\angle I $}
\newcommand{\cax}{$\angle I_c $}
\newcommand{\cix}{$I^{\text{im}}_c$}
\newcommand{\crx}{$I^{\text{re}}_c$}
\newcommand{\cmx}{$|I|_c$}
\newcommand{\px}{$P_c$}
\newcommand{\qx}{$Q_c$}
\newcommand{\w}{$W$}

\newcommand{\cixm}{I^{\text{im}}_c}
\newcommand{\crxm}{I^{\text{re}}_c}
\newcommand{\cmxm}{|I|_c}
\newcommand{\vim}{U^{\text{im}}}
\newcommand{\vrm}{U^{\text{re}}}
\newcommand{\cam}{\angle I}
\newcommand{\caxm}{\angle I_c}
\newcommand{\vmm}{|U|}
\newcommand{\vam}{\angle U}
\newcommand{\pxm}{P_c}
\newcommand{\qxm}{Q_c}
\newcommand{\wm}{W}

\newcommand{\N}{\textbf{\textcolor{blue}{N}}}

\newcommand{\varspace}{$\mathcal{X}$}
\newcommand{\varspacem}{\mathcal{X}}

State estimation determines the \textit{most-likely} state of a network, i.e., assigns numerical values to a network’s variable space \varspace, given a set of measurements \meas. Let the set~$\mathcal{X}$ include all network variables such as voltage, current, power, etc. Some variables can be expressed in multiple fashions, e.g., complex voltage phasors can either be described in polar or rectangular form. Each measurement $m \in \mathcal{M}$ is associated to an element of the variable space~$x_m \in \mathcal{X}$, its measured value $z_m$ and its standard deviation $\sigma_m$. Different power flow formulations are characterized by different variable spaces~$\mathcal{X}^{\text{form}} \subset \mathcal{X}$.

In the "standard" static SE approach, the measurements are collected in a vector \z, and $\mathbf{x} \in \varspacem^{\text{form}}$ are typically the voltage phasors at every bus: $\mathbf{U}~:=~[\mathbf{U}_i]_{i \in \busesm}$, where \buses \ is the set of all network buses. The relationship between \z \ and $\mathbf{x}$ is defined by \h, a set of (in general) nonlinear power flow equations. In practice, measurements are affected by errors, that are represented as a vector \errv, such that:

\begin{equation}\label{eq:the_se_equation}
    \boldsymbol{\eta} = \mathbf{z} - \mathbf{h}(\mathbf{x}).
\end{equation}

A standard WLS SE "filters" the measurement errors to retrieve the \textit{most-likely} state~\cite{bible}, by minimizing:

\begin{equation}\label{eq:J}
    J(\mathbf{x}) = \left[\mathbf{z} - \mathbf{h}(\mathbf{x}) \right]^T \boldsymbol{\Sigma}^{-1} \left[\mathbf{z} - \mathbf{h}(\mathbf{x}) \right],
\end{equation}

where $\boldsymbol{\Sigma}$ is a $m \times m$ diagonal matrix that contains the weights of each measurement. Under the typical assumption that measurement errors follow univariate gaussian distributions, the weights are the inverse of the measurements' variances: $\boldsymbol{\Sigma} = \text{diag} \{ \sigma_1^2, ..., \sigma_m^2 \}$. The lower the variance, the higher the confidence on the measurement accuracy. Eq.~\eqref{eq:J} is typically solved iteratively with a Gauss-Newton method~\cite{bible}.

In this work, univariate gaussian error assumptions are kept, while the SE problem is reformulated as an optimal power flow (OPF) problem, modelled and solved using mathematical optimization toolboxes. The general problem outline is: 

\begin{IEEEeqnarray}{lCr}
 & \text{minimize} \; \; \sum_{\substack{m \in \measm}}   \rho_m, \label{eq:objective} \\
  &\text{subject to:} \nonumber \\
    & \mathbf{f}(\boldsymbol{\rho}, \mathbf{x}) = 0, \label{eq:f} \\
    & \mathbf{h}(\mathbf{x}) = 0, \label{eq:h} \\    
    &  \mathbf{g}(\mathbf{x}) = 0, \label{eq:g} \\
    &  \mathbf{k}(\mathbf{x}) \leq 0. \label{eq:k} 
\end{IEEEeqnarray}

Eq.~\eqref{eq:f} represent the definition of the residuals $\boldsymbol{\rho}_m~:=~\left[ \rho_{m} \right], \; m \in \measm$, which incorporate the errors $\boldsymbol{\eta}$, and its form is driven by the used SE criterion, e.g., WLS or WLAV. The extended mathematical formulation of~\eqref{eq:f} can be found in Section~\ref{sec:residual_definition}. The objective~\eqref{eq:objective} minimizes the sum of residuals. 

Equations~\eqref{eq:h} represent the standard equality constraints of an OPF problem~\cite{PMD_PSCC}, describing Ohm's and Kirchhoff's laws. Their mathematical expression depends on the chosen power flow formulation. If all measured quantities pertain to the space of the chosen formulation $\varspacem^{\text{form}}$, eq.~\eqref{eq:objective}-\eqref{eq:h} fully describe the SE problem. If $x_m$ is not in $\varspacem^{\text{form}}$,~\eqref{eq:g} is used to couple those measurement variables \varspace$^{\text{form}}$. These are mapping functions: $g_m: \varspacem \mapsto \varspacem^{\text{form}}$, and their expressions can be found in Section~\ref{sec:measurements_from_a_different_variable_space}. Delegating the mapping functionality to $\mathbf{g} $, instead of integrating it in \h \ like in the standard SE approach \eqref{eq:the_se_equation}, allows to get a better overview of the computational complexity associated to each $(x_m, \varspacem)$ combination. This helps to choose the appropriate SE formulation given the measurement devices available in a real-life situation.

It can be observed that formulating SE in optimization terms allows for a more general problem description, which "encapsulates" different SE formulations from the literature, including those that present inequality constraints. 
The use of OPF constraints \eqref{eq:h}, and variable spaces larger than the standard: $\mathbf{U} \subset \varspacem^{\text{form}}$, avoids the need to explicitly introduce virtual measurements, as they are naturally included in \eqref{eq:h}. Furthermore, it is not necessary to add a separate $h_m$ \ for every different measured quantity, as long as $x_m \in \varspacem^{\text{form}}$.

Inequality constraints on maximum/minimum voltage, current and power values, typical of OPF, are normally not part of SE problems. With the proposed approach, these can optionally be included \eqref{eq:k}, either to accommodate specific purposes or to reduce the solver's search space. The latter can reduce computation time, but the chosen variable bounds should be sensible, as to not cut the feasible space.

\subsection{Definition of residuals}\label{sec:residual_definition}

The residual definitions presented in this work can be written as a generic p-norm: 
\begin{equation}\label{eq:p-norm} 
    \rho_m = \| x_m - z_m \|_p/\sigma^p_m \;\;, \forall m \in \measm.
\end{equation}

The SE criterion is selected through the value of $p$. Furthermore, in an optimization context, it is possible to relax \eqref{eq:p-norm}. In this paper, two residual definitions and their relaxations are addressed:~\eqref{eq:p-norm} with $p = 2$, which returns the standard WLS minimization, and~\eqref{eq:p-norm} with $p = 1$, which results in minimizing WLAV.

\subsubsection{WLS}

Eq. \eqref{eq:p-norm} with $p = 2$ is equivalent to:
\begin{equation}\label{eq:WLS}
\rho_m =  (x_m - z_m)^2/\sigma^2_m, \forall m \in \mathcal{M}.
\end{equation}

Eq.~\eqref{eq:WLS} is quadratic. To reduce computational complexity, a relaxation of~\eqref{eq:WLS} can be performed, by replacing the equality sign in~\eqref{eq:p-norm} with a $\geq$. In this paper, the relaxation is named "rWLS" and its performance is analyzed. 

\subsubsection{WLAV}

Eq. \eqref{eq:p-norm} with $p = 1$ is equivalent to:
\begin{equation}\label{eq:WLAV}
\rho_m =  | x_m - z_m |/\sigma_m, \forall m \in \mathcal{M}.
\end{equation}

Solving the WLAV minimization with a standard Gauss-Newton approach is nontrivial, as the absolute value function is not continuously differentiable. This can also cause problems to off-the-shelf solvers, but in an optimization context the residual definition in~\eqref{eq:WLAV} can be replaced by its linear "rWLAV" relaxation:

\begin{equation}\label{eq:rWLAV1}
     \rho_{m} \geq \frac{x_m - z_m}{\sigma_m}, \; \; \; \;  \forall m \in \measm,
\end{equation}
\begin{equation}\label{eq:rWLAV2}
     \rho_{m} \geq - \frac{x_m - z_m}{\sigma_m} \; \; \; \;  \forall m \in \measm,
\end{equation}
which is an exact~\cite{Irving} relaxation if the problem is a minimization, as is this case.

Given the "block" structure of the presented optimization problem, it is rather easy to replace the residual definition and explore additional SE criteria. This is left for future work.

\subsection{Measurements outside the variable space}\label{sec:measurements_from_a_different_variable_space}

This section illustrates the relationship between \varspace$^{\text{form}}$ \ and the expressions of $g_m$. An overview of the different combinations available in the software tool is provided in Table~\ref{tab:variables_and_maps}. The mapping functions have been implemented in the most computationally simple form, and could be easily relaxed or approximated in future extensions.

The formulations used in this paper are the "AC" formulation~\cite{Mahdad}, with power and voltage variables in polar (ACP) or rectangular (ACR) form, the IVR formulation~\cite{GethIVR}, with current and voltage rectangular variables and the LinDist3Flow formulation~\cite{Sankur}, with power and lifted voltage variables. A substantial part of the implementation of these formulation is imported from PowerModelsDistributions.jl~\cite{PMD_PSCC}. Relying on PowerModelsDistribution.jl allows to easily extend the present work with the other formulations available in the package. Here, the choice fell on the ACP, ACR, IVR and LinDist3Flow because the first three are exact, i.e., no modelling error is added to the SE routine, while the LinDist3Flow is a linearization, and therefore introduces modeling errors, but it also presents fast convergence and convergence guarantees. The \varspace$^{\text{form}}$ \ for each formulation is highlighted with the symbol \N \ in Table~\ref{tab:variables_and_maps}, and the variables are represented as follows: 

\begin{IEEEdescription}[\IEEEusemathlabelsep\IEEEsetlabelwidth{xxxxxxxxxxx}]
	\item[\px, \qx] Active, reactive power, 
	\item[\va, \cax] Voltage, current angle,
	\item[\vm, \cmx] Voltage, current magnitude,
	\item[\w = \vm$^2$] Lifted voltage variable,
	\item[\vr, \crx] Voltage, current - real part,
	\item[\vi, \cix] Voltage, current - imaginary part.
\end{IEEEdescription} 

The $c$ subscript indicates that the variable can refer to both the power/current flow in a branch and the power/current injection from any component $c \in \mathcal{C}$, e.g. generators, loads. The voltage is always a property of the network buses.

\begin{table*}[t]
\centering
\caption{Expression of $g_{m}$ depending on $x_m$ and power flow formulation. \N: not required ($x_m \in \varspacem^{\text{form}}$), -: conversion not provided, M: conversion of type "Multiplication", F: conversion of type "Fraction", MF: conversion of type "Multiplication Fraction", PP: conversion of type "Tangent", S: conversion of type "Square", SF: conversion of type "Square Fraction".}
\begin{tabular}{| l||c|c|c|c|c|c|c|c|c|c|c|}
 \hline
 \multicolumn{12}{|c|}{Expression of the mapping function $g_{m}: \varspacem \mapsto \varspacem^{\text{form}}$} \\
 \hline
  \multicolumn{1}{|c||}{} &\multicolumn{11}{c|}{$x_m$} \\ \hline
 Formulation & \vm & \va & \vr & \vi & \w & \px & \qx & \cmx & \ca & \crx & \cix \\
 \hline
ACP          & \N & \N  & - & - & - & \N    & \N     & SF & -  & F  & F \\ \hline
ACR          & S & T & \N & \N & - & \N    & \N     & SF & -  & MF & MF \\ \hline
IVR          & S & T & \N & \N & - & M    & M     & S  & T & \N  & \N \\ \hline
LinDist3Flow & S & -  & - & - & \N & \N    & \N     & SF & -  & -  & - \\

 \hline
\end{tabular}\label{tab:variables_and_maps}
\end{table*}


\subsubsection{Tangent}

The conversion type Tangent allows to include \va \ measurements in the ACR and IVR formulation, and \cax \ measurements in the IVR formulation, respectively through:

\begin{eqnarray}
     & \tan(\vam) = \vim/\vrm, \\ 
     & \tan(\caxm) = \cixm/\crxm,
\end{eqnarray}

which are nonlinear trigonometric constraints. 

\subsubsection{Fraction}
The conversion type Fraction allows to include \crx \ and \cix \ measurements in the ACP formulation, respectively through:

\begin{eqnarray}
    &  \crxm = \dfrac{\pxm \cdot \cos(\vam)+\qxm \cdot \sin(\vam)}{\vmm}, \\
    &  \cixm = \dfrac{\pxm \cdot \sin(\vam)-\qxm \cdot \cos(\vam)}{\vmm},
\end{eqnarray}

which are nonlinear trigonometric constraints.

\subsubsection{Multiplication}
The conversion type Multiplication allows to include \px \ and \qx \ measurements in the IVR formulation, respectively through:
\begin{eqnarray}\label{eq:multiplication}
     & \pxm = \vrm \cdot \crxm + \vim \cdot \cixm, \\
     & \qxm = \vim \cdot \crxm - \vrm \cdot \cixm,
\end{eqnarray}

which are quadratic constraints. This makes them more tractable than generic non-convex constraints and algebraic modeling framework such as JuMP do not need to perform automatic differentiation, as derivative information on quadratic forms is known.

\subsubsection{Multiplication Fraction} 
The conversion type Multiplication Fraction allows to include \crx \ and \cix \ measurements in the ACR formulation, respectively through:
\begin{eqnarray}
    & \crxm = \dfrac{\pxm \cdot \vrm + \qxm \cdot \vim}{(\vrm)^2 + (\vim)^2}, \\
    & \cixm = \dfrac{\pxm \cdot \vim - \qxm \cdot \vrm}{(\vrm)^2 + (\vim)^2},
\end{eqnarray}
 which are non-convex constraints.

\subsubsection{Square Fraction}

The conversion type Square Fraction allows to include \cmx \ measurements in the ACP and ACR formulation, through:

\begin{equation}
      (\cmxm)^2  = \frac{ (\pxm)^2 + (\qxm)^2 }{(\vmm)^2}  
\end{equation}

which is a non-convex constraint.
If the conversion is applied to the LinDist3Flow, then \vm$^2$ is replaced by \w. 

\subsubsection{Square}

The "Square" allows to include \vm \ measurements in the ACR and IVR formulation, and \cmx \ measurements in the IVR formulation, respectively through:

\begin{eqnarray}\label{eq:square}
    &  (\vmm)^2  = (\vim)^2 + (\vrm)^2,                 \\
    &  (\cmxm)^2 = (\cixm)^2 + (\crxm)^2,   
\end{eqnarray}

which are quadratic constraints.

\subsubsection{Not provided conversions}

As can be observed in Table~\ref{tab:variables_and_maps}, some conversions are not provided. This is because the measured quantities are either unlikely to take place in practice, e.g., \w, or tend to appear in pairs, e.g., with PMUs, \cax \ is measured together with \cmx. In the latter case, it is more efficient to transform \cax, \cmx \ in \crx, \cix \ and then use these measurements, as they are either part of the IVR space or the conversions starting from \crx, \cix \ are computationally less complex than those from \cax, \cmx.




%
\section{Case studies}\label{sec:case_studies}
In this section, four case studies illustrate possible applications of the framework. Section~\ref{sec:comparison_criteria} illustrates that the SE errors and the solve times with off-the shelf solvers are reasonable, and thus the implementation is reliable. Furthermore, it showcases the ease of comparison between different exact formulations and estimation criteria. Similarly, Section~\ref{sec:comparison_formulation} presents a comparison between the fastest exact formulation and the LinDist3Flow, illustrating that the framework can help determine whether the trade-off between decreased complexity and presence of modeling errors make a linear formulation appealing. Section~\ref{sec:comparison_measurements} shows that the framework can deal with different measurement devices and allows to exploit the advantages of PMUs from a computational standpoint. Finally, Section~\ref{sec:underdetermined system} shows how the SE error varies depending on the amount of monitored houses, displaying the possibility of handling underdetermined systems without the need of adding pseudo measurements.

The first three case studies assume that the system is fully monitored, i.e., there are no pseudo measurements, and that there are no bad data. These assumptions are not realistic for distribution systems, but they allow to prove that the framework works, without adding interpretability problems due to data quality issues. Addressing realistic scenarios is the next step and is left for future work.

To show that the framework is robust and scalable, simulations are performed on the ENWL database \cite{ENWL}: a collection of 25 real-life unbalanced LV networks (128 feeders in total) and power profiles. 
The data, scripts and results of the simulations performed in this paper can be found online \cite{scripts_repo} for benchmarking and reproducibility. The calculations are run on a 64-bit machine with Intel(R) Xeon(R) CPU E5-4610 v4 @1.80GHz and 32 GB RAM, using julia 1.5.2, PowerModelsDistribution.jl 0.9.1, PowerModelsSE.jl 0.1.2. For nonlinear problems, the solver used is Ipopt 3.12.10~\cite{ipopt}, with Ipopt.jl v0.6.2 and the HSL MA27 subroutine~\cite{HSL}. For problems with linear constraints, Gurobi 9.0.3 with Gurobi.jl 0.8.1 is used.

As often done in practice, weights $\sigma$ assigned to the measurements correspond to one third of the maximum measurement error of the relative measuring devices~\cite{Muscas}. The meter accuracy is typically subject to government regulation, as the readings are used for billing purposes. The accuracy is typically in the class of 0.5\% ~\cite{Irwin}. 
Therefore, the $\sigma_m$ for both smart meters and PMUs is assumed to be 1.67\textperthousand. Furthermore, voltage magnitude and total feeder demand are measured at the substation, with a higher precision device of $\sigma_m = 0.33$\textperthousand. 

To create measurement data, power flow calculations are performed on each feeder, and errors are added to the solution in correspondence of every meter, sampling from its Gaussian distribution. These are then used as input for state estimation calculations. It is assumed that all customers have a smart meter that measures \vm, $P$ and  $Q$. Additionally, \vm, $P, \ Q$ are also measured at the substation, for each feeder. The voltage angles at the substation are fixed~\cite{bible}.


Each feeder's maximum ($\varepsilon^{max}$) and average ($\varepsilon^{avg}$) absolute state estimation error are presented, where the error $\varepsilon$ is calculated for the voltage magnitude at each bus:
\begin{equation}
    \varepsilon_i = | \vmm^{pf}_i -\vmm^{se}_i| \; \; \; \; \forall i \in \busesm,
\end{equation}

where $\vmm^{pf}$ is from the solution of the power flow and $\vmm^{se}$ of the state estimation.

The displayed SE results refer to a single time stamp: consumer power profiles are extracted from the ENWL database for a summer day at 12PM, and include the presence of PV.

\subsection{Comparison of SE criteria with exact formulations}\label{sec:comparison_criteria}

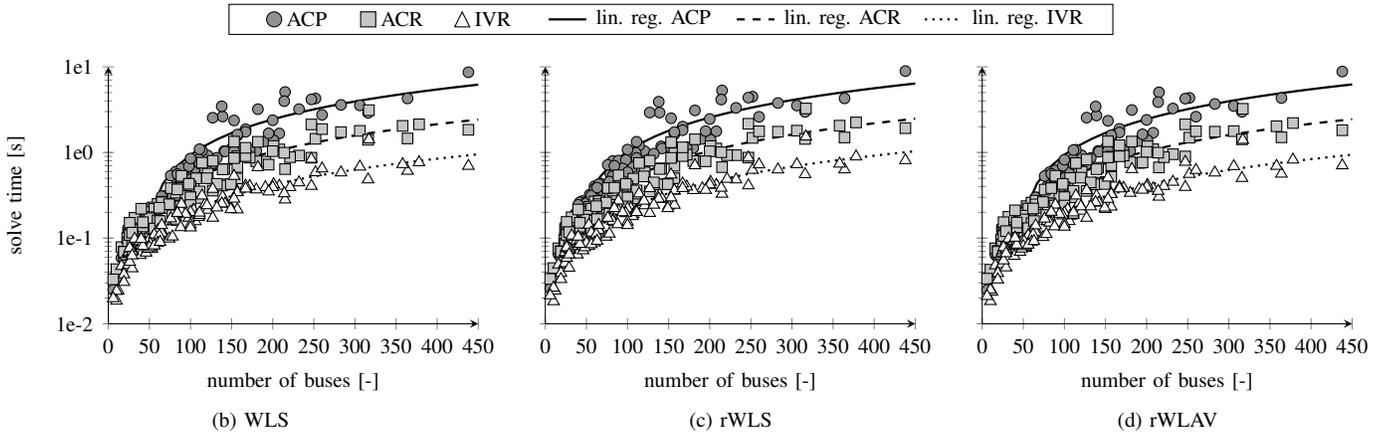
\begin{figure*}[h]
    \centering
    \subfloat
    {
    \centering
    \begin{tikzpicture}
        \draw (0.0,0.05) rectangle (11.75,0.45);
        \draw [fill=gray!85] (0.6,0.25) circle (0.1cm);
        \draw (0.5,0.25) node [right] {\footnotesize ~\,ACP};
        \draw [fill=gray!45] (1.75,0.15) rectangle (1.95,0.35);
        \draw (1.75,0.25) node [right] {\footnotesize ~\,ACR};
        \draw [fill=gray!5] (3.0,0.15) -- (3.2,0.15) -- (3.1,0.35) -- (3.0,0.15);
        \draw (3.0,0.25) node [right] {\footnotesize ~\,IVR};
        \draw [line width = 0.3mm](4.25,0.25) -- (4.75,0.25) node [right] {\footnotesize lin. reg. ACP};
        \draw [dashed, line width = 0.3mm] (6.75,0.25) -- (7.25,0.25) node [right] {\footnotesize lin. reg. ACR};
        \draw [dotted, line width = 0.3mm] (9.25,0.25) -- (9.75,0.25) node [right] {\footnotesize lin. reg. IVR};
    \end{tikzpicture}
    }
    \vspace{-0.5cm}
    \vskip\baselineskip
    \subfloat[WLS]
    {
	\begin{tikzpicture}
	    \begin{semilogyaxis}[ 	width = 6.5cm, height = 5.00cm, 
		    		            xlabel={\footnotesize number of buses [-]}, 
		    		            ylabel={\footnotesize solve time [s]},
		    		            xmin=0, xmax=450, ymin=0.01, ymax=10,
		    		            xtick = {0, 50, 100, 150, 200, 250, 300, 350, 400, 450}, 
		    		            xticklabels = {\footnotesize 0, \footnotesize 50, \footnotesize 100, \footnotesize 150, \footnotesize 200, \footnotesize 250, \footnotesize 300, \footnotesize 350, \footnotesize 400, \footnotesize 450},
		    		            ytick = {0.01, 0.1, 1, 10}, 
		    		            yticklabels = {\footnotesize 1e-2, \footnotesize 1e-1, \footnotesize 1e0, \footnotesize 1e1}, 
		    		            axis lines=left, scaled y ticks = false, axis on top, clip=true]
		    \addplot [every mark/.append style={solid, fill=gray!85}, mark=*, mark size=2.0pt, only marks]
		        table [x expr={and(\strequal{\thisrow{criterion}}{wls},\strequal{\thisrow{eq_model}}{rACP})==1?\thisrow{n_bus}:nan}, y=solve_time, col sep=comma] {csv/case_study_1_clean.csv};
		    \addplot [every mark/.append style={solid, fill=gray!45}, mark=square*, mark size=2.0pt, only marks]
		        table [x expr={and(\strequal{\thisrow{criterion}}{wls},\strequal{\thisrow{eq_model}}{rACR})==1?\thisrow{n_bus}:nan}, y=solve_time, col sep=comma] {csv/case_study_1_clean.csv};
		    \addplot [every mark/.append style={solid, fill=gray!5}, mark=triangle*, mark size=2.5pt, only marks]
		        table [x expr={and(\strequal{\thisrow{criterion}}{wls},\strequal{\thisrow{eq_model}}{rIVR})==1?\thisrow{n_bus}:nan}, y=solve_time, col sep=comma] {csv/case_study_1_clean.csv};
		    \addplot [domain=50:450, line width = 0.3mm] expression {0.0150199*x-0.566041}; 
		    \addplot [domain=50:450, dashed, line width = 0.3mm] expression {0.00559247*x-0.0983817}; 
		    \addplot [domain=50:450, dotted, line width = 0.3mm] expression {0.00213573*x-0.00930781}; 
		\end{semilogyaxis}
	\end{tikzpicture}
	}
    \subfloat[rWLS]
	{
	\begin{tikzpicture}
	    \begin{semilogyaxis}[ 	width = 6.5cm, height = 5.00cm, 
		    		            xlabel={\footnotesize number of buses [-]},
		    		            xmin=0, xmax=450, ymin=0.01, ymax=10,
		    		            xtick = {0, 50, 100, 150, 200, 250, 300, 350, 400, 450}, 
		    		            xticklabels = {\footnotesize 0, \footnotesize 50, \footnotesize 100, \footnotesize 150, \footnotesize 200, \footnotesize 250, \footnotesize 300, \footnotesize 350, \footnotesize 400, \footnotesize 450},
		    		            ytick = {0.01, 0.1, 1, 10}, 
		    		            yticklabels = {\footnotesize , \footnotesize , \footnotesize , \footnotesize }, 
		    		            axis lines=left, scaled y ticks = false, axis on top, clip=true]
		    \addplot [every mark/.append style={solid, fill=gray!85}, mark=*, mark size=2.0pt, only marks] 
		        table [x expr={and(\strequal{\thisrow{criterion}}{rwls},\strequal{\thisrow{eq_model}}{rACP})==1?\thisrow{n_bus}:nan}, y=solve_time, col sep=comma] {csv/case_study_1_clean.csv};
		    \addplot [every mark/.append style={solid, fill=gray!45}, mark=square*, mark size=2.0pt, only marks]
		        table [x expr={and(\strequal{\thisrow{criterion}}{rwls},\strequal{\thisrow{eq_model}}{rACR})==1?\thisrow{n_bus}:nan}, y=solve_time, col sep=comma] {csv/case_study_1_clean.csv};
		    \addplot [every mark/.append style={solid, fill=gray!5}, mark=triangle*, mark size=2.5pt, only marks]
		        table [x expr={and(\strequal{\thisrow{criterion}}{rwls},\strequal{\thisrow{eq_model}}{rIVR})==1?\thisrow{n_bus}:nan}, y=solve_time, col sep=comma] {csv/case_study_1_clean.csv};
		    \addplot [domain=50:450, line width = 0.3mm] expression {0.015298*x-0.499883}; 
		    \addplot [domain=50:450, dashed, line width = 0.3mm] expression {0.00573807*x-0.105198}; 
		    \addplot [domain=50:450, dotted, line width = 0.3mm] expression {0.00232937*x-0.0148941}; 
	    \end{semilogyaxis}
    \end{tikzpicture}
    }
    \subfloat[rWLAV]
    {
	\begin{tikzpicture}
	    \begin{semilogyaxis}[ 	width = 6.5cm, height = 5.00cm, 
		    		            xlabel={\footnotesize number of buses [-]}, 
		    		            xmin=0, xmax=450, ymin=0.01, ymax=10,
		    		            xtick = {0, 50, 100, 150, 200, 250, 300, 350, 400, 450}, 
		    		            xticklabels = {\footnotesize 0, \footnotesize 50, \footnotesize 100, \footnotesize 150, \footnotesize 200, \footnotesize 250, \footnotesize 300, \footnotesize 350, \footnotesize 400, \footnotesize 450},
		    		            ytick = {0.01, 0.1, 1, 10}, 
		    		            yticklabels = {\footnotesize , \footnotesize , \footnotesize , \footnotesize }, 
		    		            axis lines=left, scaled y ticks = false, axis on top, clip=true]
		    \addplot [every mark/.append style={solid, fill=gray!85}, mark=*, mark size=2.0pt, only marks] 
		        table [x expr={and(\strequal{\thisrow{criterion}}{rwlav},\strequal{\thisrow{eq_model}}{rACP})==1?\thisrow{n_bus}:nan}, y=solve_time, col sep=comma] {csv/case_study_1_clean.csv};
		    \addplot [every mark/.append style={solid, fill=gray!45}, mark=square*, mark size=2.0pt, only marks] 
		        table [x expr={and(\strequal{\thisrow{criterion}}{rwlav},\strequal{\thisrow{eq_model}}{rACR})==1?\thisrow{n_bus}:nan}, y=solve_time, col sep=comma] {csv/case_study_1_clean.csv};
		    \addplot [every mark/.append style={solid, fill=gray!5}, mark=triangle*, mark size=2.5pt, only marks] 
		        table [x expr={and(\strequal{\thisrow{criterion}}{rwlav},\strequal{\thisrow{eq_model}}{rIVR})==1?\thisrow{n_bus}:nan}, y=solve_time, col sep=comma] {csv/case_study_1_clean.csv};
		    \addplot [domain=50:450, line width = 0.3mm] expression {0.0151183*x-0.572568}; 
		    \addplot [domain=50:450, dashed, line width = 0.3mm] expression {0.00565655*x-0.0998161}; 
		    \addplot [domain=50:450, dotted, line width = 0.3mm] expression {0.00210785*x-0.00733968}; 
	    \end{semilogyaxis}
    \end{tikzpicture}
    }    
    \caption{Solve time for different formulations and criteria.}
    \label{fig:time_comparison_cs1}
\end{figure*}

Fig.\ref{fig:time_comparison_cs1} shows the solve time for the first case study. In all scatter plots, a data point corresponds to a feeder. It can be observed that the computational difference amongst the estimation criteria is modest, whereas, the SE based on the IVR formulation largely outperforms the ACR and ACP, despite the need to include the nonlinear \eqref{eq:multiplication}, \eqref{eq:square} to incorporate voltage and power measurements from smart meters. 

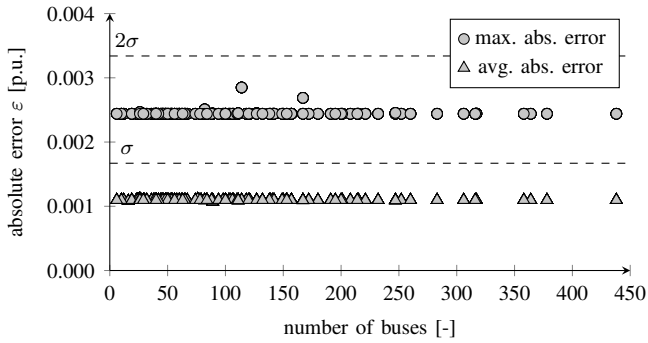
\begin{figure}[h]
    \centering
    \begin{tikzpicture}
        \begin{axis}[ 	width = 8.5cm, height = 5.00cm, 
			    	    xlabel={\footnotesize number of buses [-]}, 
			    		ylabel={\footnotesize absolute error~$\varepsilon$ [p.u.]},
			    		xmin=0, xmax=450, ymin=0.0, ymax=0.004,
			    		xtick = {0, 50, 100, 150, 200, 250, 300, 350, 400, 450}, 
			    		xticklabels = {\footnotesize 0, \footnotesize 50, \footnotesize 100, \footnotesize 150, \footnotesize 200, \footnotesize 250, \footnotesize 300, \footnotesize 350, \footnotesize 400, \footnotesize 450},
			    		ytick = {0.0, 0.001, 0.002, 0.003, 0.004}, 
			    		yticklabels = {\footnotesize 0.000, \footnotesize 0.001, \footnotesize 0.002, \footnotesize 0.003, \footnotesize 0.004}, 
			    		axis lines=left, scaled y ticks = false, axis on top, clip=true]
	        \addplot [every mark/.append style={solid, fill=gray!45}, mark=*, mark size=2.0pt, only marks] 
			        table [x = n_bus, y=err_max, col sep=comma] {csv/case_study_1_clean.csv};
			\addlegendentry{\footnotesize max. abs. error}
			\addplot [every mark/.append style={solid, fill=gray!45}, mark=triangle*, mark size=2.5pt, only marks] 
			        table [x = n_bus, y=err_avg, col sep=comma] {csv/case_study_1_clean.csv};
			\addlegendentry{\footnotesize avg. abs. error}
			\addplot [dashed] coordinates {(0.0,0.00167) (450.0,0.00167)}; 
			\node at (axis cs:15.0,0.00167) [above] {\footnotesize $\sigma$};
			\addplot [dashed] coordinates {(0.0,0.00334) (450.0,0.00334)};
			\node at (axis cs:15.0,0.00334) [above] {\footnotesize $2\sigma$};
	    \end{axis}
    \end{tikzpicture}
    \caption{Absolute error for any exact formulation and criterion.}
    \label{fig:errors_cs1}
\end{figure}

The three analysed criteria present the same accuracy (same $\varepsilon^{max}, \varepsilon^{avg}$) in all feeders, which means that the implementation is consistent and reliable. The absolute errors are displayed in Fig. \ref{fig:errors_cs1}. The $\varepsilon^{max}$ never exceeds 0.003 p.u., which is less than 0.70V in a 230V feeder, except for one outlier, removed from the plot for ease of visualization. The outlier error is extremely high: $\varepsilon^{avg} \approx 0.023 $ and $\varepsilon^{max} \approx 0.083$, and occurs for feeder 4 in network 13. A dedicated analysis showed that this feeder presents an average voltage unbalance factor above 6\%. Therefore, it is an ill-conditioned case that largely exceeds the acceptable operational limits, so faulty SE seems justifiable.

In conclusion, all the exact formulations and criteria present the same, modest, errors, which confirm that the SE implementation is reliable. The clear computational advantage of the IVR indicates that the choice of the formulation can have a significant impact on the SE performance.

\subsection{Comparison of linearized and exact formulations}\label{sec:comparison_formulation}

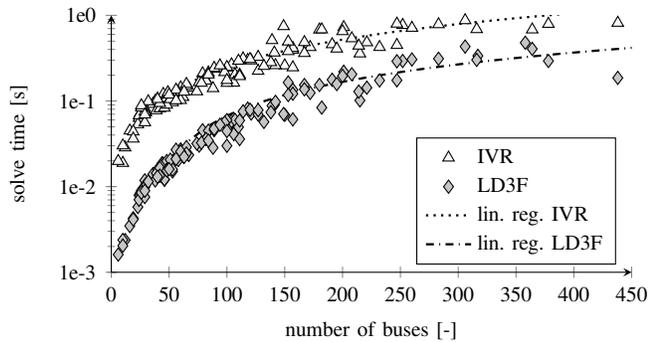
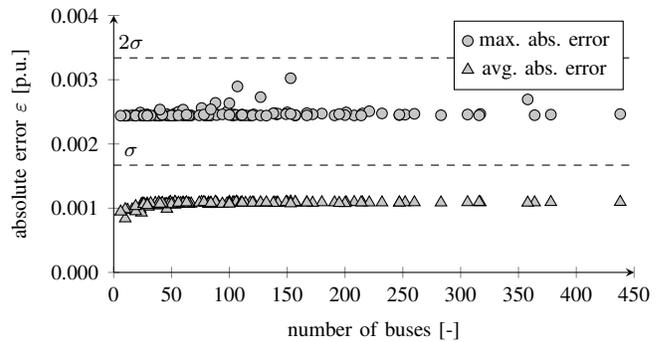
\begin{figure*}[h]
    \centering
    \subfloat[Solve time for IVR and LD3F formulation and rWLAV criterion.] 
	{
	\begin{tikzpicture}
	    \begin{semilogyaxis}[ 	width = 8.5cm, height = 5.00cm, 
			    		        xlabel={\footnotesize number of buses [-]}, 
			    		        ylabel={\footnotesize solve time [s]},
			    		        xmin=0, xmax=450, ymin=0.001, ymax=1,
			    		        xtick = {0, 50, 100, 150, 200, 250, 300, 350, 400, 450}, 
			    		        xticklabels = {\footnotesize 0, \footnotesize 50, \footnotesize 100, \footnotesize 150, \footnotesize 200, \footnotesize 250, \footnotesize 300, \footnotesize 350, \footnotesize 400, \footnotesize 450},
			    		        ytick = {0.001, 0.01, 0.1, 1}, 
			    		        yticklabels = {\footnotesize 1e-3, \footnotesize 1e-2, \footnotesize 1e-1, \footnotesize 1e0}, 
			    		        axis lines=left, scaled y ticks = false, axis on top, clip=true,
			    		        legend cell align={left},legend pos=south east]
		\addplot [every mark/.append style={solid, fill=gray!5}, mark=triangle*, mark size=2.5pt, only marks]
			 table [x expr={and(\strequal{\thisrow{criterion}}{rwlav},\strequal{\thisrow{eq_model}}{rIVR})==1?\thisrow{n_bus}:nan}, y=solve_time, col sep=comma] {csv/case_study_2_clean.csv};
			    \addlegendentry{\footnotesize IVR}
			    \addplot [every mark/.append style={solid, fill=gray!45}, mark=diamond*, mark size=2.5pt, only marks]
			        table [x expr={and(\strequal{\thisrow{criterion}}{rwlav},\strequal{\thisrow{eq_model}}{LD3F})==1?\thisrow{n_bus}:nan}, y=solve_time, col sep=comma] {csv/case_study_2_clean.csv};
			    \addlegendentry{\footnotesize LD3F}
			    \addplot [domain=50:450, dotted, line width = 0.3mm] expression {0.00270291*x-0.0200099}; 
			    \addlegendentry{\footnotesize lin. reg. IVR}
			    \addplot [domain=50:450, dashdotted, line width = 0.3mm] expression {0.00100054*x-0.0327682}; 
			    \addlegendentry{\footnotesize lin. reg. LD3F}
		    \end{semilogyaxis}
	    \end{tikzpicture}
    }
    \subfloat[Absolute error for the LD3F formulation.]
    {
	\begin{tikzpicture}
        \begin{axis}[ 	width = 8.5cm, height = 5.00cm, 
			    	    xlabel={\footnotesize number of buses [-]}, 
			    		ylabel={\footnotesize absolute error~$\varepsilon$ [p.u.]},
			    		xmin=0, xmax=450, ymin=0.0, ymax=0.004,
			    		xtick = {0, 50, 100, 150, 200, 250, 300, 350, 400, 450}, 
			    		xticklabels = {\footnotesize 0, \footnotesize 50, \footnotesize 100, \footnotesize 150, \footnotesize 200, \footnotesize 250, \footnotesize 300, \footnotesize 350, \footnotesize 400, \footnotesize 450},
			    		ytick = {0.0, 0.001, 0.002, 0.003, 0.004}, 
			    		yticklabels = {\footnotesize 0.000, \footnotesize 0.001, \footnotesize 0.002, \footnotesize 0.003, \footnotesize 0.004}, 
			    		axis lines=left, scaled y ticks = false, axis on top, clip=true]
	        \addplot [every mark/.append style={solid, fill=gray!45}, mark=*, mark size=2.0pt, only marks] 
			        table [x expr={and(\strequal{\thisrow{criterion}}{rwlav},\strequal{\thisrow{eq_model}}{LD3F})==1?\thisrow{n_bus}:nan}, y=err_max, col sep=comma] {csv/case_study_2_clean.csv};
			\addlegendentry{\footnotesize max. abs. error}
			\addplot [every mark/.append style={solid, fill=gray!45}, mark=triangle*, mark size=2.5pt, only marks] 
			        table [x expr={and(\strequal{\thisrow{criterion}}{rwlav},\strequal{\thisrow{eq_model}}{LD3F})==1?\thisrow{n_bus}:nan}, y=err_avg, col sep=comma] {csv/case_study_2_clean.csv};
			\addlegendentry{\footnotesize avg. abs. error}
			\addplot [dashed] coordinates {(0.0,0.00167) (450.0,0.00167)}; 
			\node at (axis cs:15.0,0.00167) [above] {\footnotesize $\sigma$};
			\addplot [dashed] coordinates {(0.0,0.00334) (450.0,0.00334)};
			\node at (axis cs:15.0,0.00334) [above] {\footnotesize $2\sigma$};
	    \end{axis}
    \end{tikzpicture}
    }
    \caption{Results for the second case study: comparison of the fastest exact formulation (IVR) and the LD3F linearization.}
    \label{fig:case_study_2}
\end{figure*}

In this case study, the most performant exact SE (IVR) is compared to SE with the LD3F formulation. The latter is a linearization based on the assumptions that line losses are limited and phases are almost balanced~\cite{Sankur}, which therefore introduces modeling errors in \eqref{eq:h}. To avoid nonlinear constraints, the \vm \ measurements are squared so they correspond to the lifted $W$. Again, the difference between the different criteria has proven negligible. Therefore, only the results using the rWLAV criteria are shown (Fig.~\ref{fig:case_study_2}). With respect to the first case study, a different Monte Carlo sample for the meters errors has been taken for the IVR, to observe whether there are significant different in the solve time.

Fig.~\ref{fig:case_study_2}(a) shows that as expected, being fully linear, the LD3F is (on average 72\%) faster than the IVR. Moreover, Fig.~\ref{fig:case_study_2}(b) illustrates that the SE accuracy is very similar to that of the exact formulations in Fig.~\ref{fig:errors_cs1}. While a more detailed analysis is needed to assess whether this holds in a sufficient number of scenarios, the trade-off between speed and accuracy for the LD3F SE seems worth investigating.

\subsection{Comparison of linear and nonlinear IVR}\label{sec:comparison_measurements}

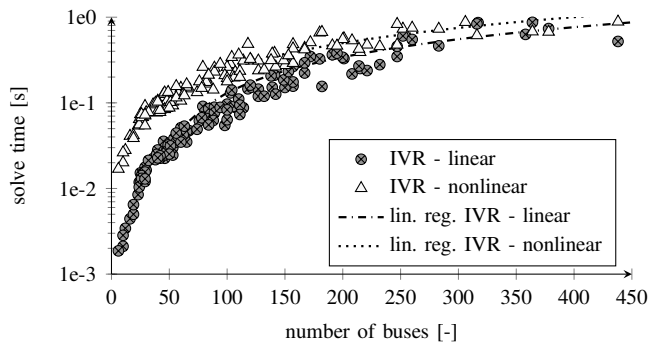
\begin{figure}[h]
    \centering
    \begin{tikzpicture}
		    \begin{semilogyaxis}[ 	width = 8.5cm, height = 5.00cm, 
			    		            xlabel={\footnotesize number of buses [-]}, 
			    		            ylabel={\footnotesize solve time [s]},
			    		            xmin=0, xmax=450, ymin=0.001, ymax=1,
			    		            xtick = {0, 50, 100, 150, 200, 250, 300, 350, 400, 450}, 
			    		            xticklabels = {\footnotesize 0, \footnotesize 50, \footnotesize 100, \footnotesize 150, \footnotesize 200, \footnotesize 250, \footnotesize 300, \footnotesize 350, \footnotesize 400, \footnotesize 450},
			    		            ytick = {0.001, 0.01, 0.1, 1}, 
			    		            yticklabels = {\footnotesize 1e-3, \footnotesize 1e-2, \footnotesize 1e-1, \footnotesize 1e0}, 
			    		            axis lines=left, scaled y ticks = false, axis on top, clip=true,
			    		            legend cell align={left},legend pos=south east]
			    \addplot [every mark/.append style={solid, fill=gray!85}, mark=otimes*, mark size=2.0pt, only marks]
			        table [x expr={and(\strequal{\thisrow{criterion}}{rwlav},\strequal{\thisrow{linear_solver}}{gurobi})==1?\thisrow{n_bus}:nan}, y=solve_time, col sep=comma] {csv/case_study_3_clean.csv};
			    \addlegendentry{\footnotesize IVR - linear}
			    \addplot [every mark/.append style={solid, fill=gray!5}, mark=triangle*, mark size=2.5pt, only marks]
			        table [x expr={and(\strequal{\thisrow{criterion}}{rwlav},\strequal{\thisrow{linear_solver}}{ma27})==1?\thisrow{n_bus}:nan}, y=solve_time, col sep=comma] {csv/case_study_3_clean.csv};
			    \addlegendentry{\footnotesize IVR - nonlinear}
			    \addplot [domain=50:450, dashdotted, line width = 0.3mm] expression {0.00211271*x-0.0803944}; 
			    \addlegendentry{\footnotesize lin. reg. IVR - linear}
			    \addplot [domain=50:450, dotted, line width = 0.3mm] expression {0.00251576 *x-0.00466615}; 
			    \addlegendentry{\footnotesize lin. reg. IVR - nonlinear}
		    \end{semilogyaxis}
	    \end{tikzpicture}
    \caption{Solve time for linear and nonlinear IVR.}
    \label{fig:time_comparison_cs3}
\end{figure}

In this case study, the IVR SE as seen previously is compared to an IVR SE where PMUs replace all smart meters. PMUs can measure \vm, \va, \cmx, and \cax. These devices can also typically present better accuracy and higher sampling resolution than smart meters, but these advantages are not considered. The objective here is to analyze the computational gains deriving from the use of PMUs: \vm, \va, \cmx, and \cax \ are converted into \vr, \vi, \crx, and \cix, and then used for SE with the IVR formulation, which is an exact model with linear constraints only. Therefore, it has the same accuracy as the previously seen nonlinear IVR, but should solve faster. Fig.~\ref{fig:time_comparison_cs3} shows that, indeed, the solution time is on average~$\approx$~46\% lower.

\subsection{Underdetermined system}\label{sec:underdetermined system}

\begin{figure}[h]
    \centering
    \begin{tikzpicture}
        \begin{semilogyaxis}[ 	width = 8.5cm, height = 5.00cm, 
			    	            xlabel={\footnotesize measured houses [-]}, 
			    		        ylabel={\footnotesize absolute error~$\varepsilon$ [p.u.]},
			    		        xmin=16, xmax=56, 
			    		        ymin=0.001, ymax=1.0,
			    		        xtick = {5, 10, 15, 20, 25, 30, 35, 40, 45, 50, 55}, 
			    		        xticklabels = {\footnotesize 5, \footnotesize 10, \footnotesize 15, \footnotesize 20, \footnotesize 25, \footnotesize 30, \footnotesize 35, \footnotesize 40, \footnotesize 45, \footnotesize 50, \footnotesize 55},
			    		        ytick = {0.001, 0.01, 0.1, 1.0}, 
			    		        yticklabels = {\footnotesize 1e-3, \footnotesize 1e-2, \footnotesize 1e-1, \footnotesize 1e0}, 
			    		        axis lines=left, scaled y ticks = false, axis on top, clip=true,
			    		        legend cell align={left},legend pos=south west]
	        \addplot [every mark/.append style={solid, fill=gray!45}, mark=*, mark size=2.0pt, only marks] 
			        table [x expr = {(\thisrow{n_meas}-3)/3}, y=err_max, col sep=comma] {csv/case_study_4_clean.csv};
			\addlegendentry{\footnotesize max. abs. error}
			\addplot [every mark/.append style={solid, fill=gray!45}, mark=triangle*, mark size=2.5pt, only marks] 
			        table [x expr = {(\thisrow{n_meas}-3)/3}, y=err_avg, col sep=comma] {csv/case_study_4_clean.csv};
			\addlegendentry{\footnotesize avg. abs. error}

	    \end{semilogyaxis}
    \end{tikzpicture}
    \caption{Absolute error for determined and underdetermined scenarios (case study 4), with exact IVR formulation and rwlav criterion.}
    \label{fig:case_study_4}
\end{figure}
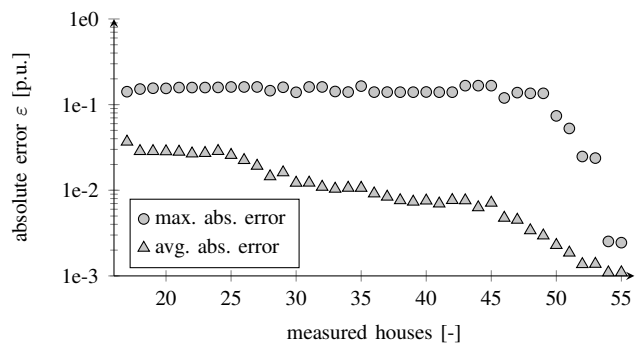

For this case study, feeder 1 of network 1 of the database is used, which corresponds to the IEEE European Low Voltage Test Feeder~\cite{ENWL}, which has 55 customers (houses). Fig.~\ref{fig:case_study_4} shows the impact of removing house meters on the SE errors. No pseudo measurements are added to replace measurements in houses without meters. This showcases the ability of the proposed framework to deal with underdetermined systems. The figure is realized by performing calculations starting with a fully monitored systems and then randomly dropping houses measurements one by one. While this is just an illustrative example, such an analysis could be useful to SE "designers" to understand the limitation of a possible implementation given the available measurement devices. However, it is clear that the SE performance deteriorates quite rapidly if no techniques are used to cope with the lack of measurements. In an optimization context, powerful techniques could be used, such as limiting the voltage drop between one house and its neighbour with an inequality constraint, in order to have a more realistic result.

\section{Conclusions}\label{sec:conclusion}
A conceptual framework has been proposed, that defines static state estimation in terms of a generalized constrained optimization problem. This generalization allows to encapsulate the multiple modelling possibilities of state estimation and compare them. Any equality or inequality constraint can be included, and underdetermined problems can be addressed. 

The conceptual framework is implemented in an open source package \cite{DSSE.jl}, that allows to flexibly combine different power flow formulations, constraints, relaxations, minimization criteria and measurement types, to obtain different state estimation models. Several case studies have been presented to showcase the functionalities of the framework/package. A takeaway from these studies is that the choice of state estimation criterion does not substantially affect the computation time, while the choice of the power flow formulation does. 

The calculations are performed on a large number of low voltage feeders to show the reliability and scalability of the approach, and data, results and scripts are also made available.

Although the concept and code can also be applied to the transmission system, the framework find its maximum usefulness in distribution networks, whose challenges and diverse features make it a harder task to design suitable state estimation routines. Facilitating the implementation of such routines has the potential to lead to a faster transition towards active distribution networks.

\bibliographystyle{IEEEtran}
\bibliography{IEEEabrv,Bibliography}

\vspace{0.2cm}
\footnotesize
\textbf{Marta Vanin} (S'19) received the M.Sc. degree in Energy Engineering from the University of Trento and the Free University of Bolzano, Italy, in 2018. She is now working towards a Ph.D. degree at KU Leuven, Belgium, in the field of computational methods for decision support to grid operators.\\
\vspace{0.2cm}\\
\textbf{Tom Van Acker} (S’14) was born in 1990, in Roeselare, Belgium. He received the M.Eng., M.Sc. and Ph.D. degrees in Electrical Engineering from KU Leuven, Belgium in 2012, 2014 and 2020, respectively. Currently, he is a post-doctoral researcher at KU Leuven and his main areas of research interest are stochastic processes and optimization in a power systems context. \\
\vspace{0.2cm}\\
\textbf{Reinhilde D'hulst} received a Master’s degree in Electrical Engineering in 2004, and obtained her PhD in the field of Power Electronics in 2009, both from KU Leuven, Belgium. Since 2009 she is with the Energy Technology department of the Flemish Institute for Technological Research (VITO) and EnergyVille, Belgium.  She is involved in several national as well as European research projects related to Smart Grids, a.o. Linear, EvolvDSO, SmartNet, EUSysflex. The main focus of her work is the development of smart grid solutions for electricity (distribution) grid-related issues, distribution grid modelling, simulation and optimisation, and flexibility assessment and control algorithms for demand response. \\
\vspace{0.2cm}\\
\textbf{Dirk Van Hertem} (S’02-SM’09) graduated as a M.Eng. in 2001 from the KHK, Geel, Belgium and as a M.Sc. in Electrical Engineering from KU Leuven, Belgium in 2003. In 2009, he has obtained his PhD, also from KU Leuven. In 2010, Dirk Van Hertem was a member of EPS group at the Royal Institute of Technology (KTH), in Stockholm. Since spring 2011 he is back at the University of Leuven where he is an associate professor in the ELECTA group. His special fields of interest are decision support for grid operators, power system operation and control in systems with FACTS and HVDC and building the transmission system of the future, including offshore grids and the supergrid concept.

\end{document}